\documentclass[sigconf]{acmart} 

\usepackage{url}
\usepackage{algorithm}
\usepackage{algpseudocode}
\usepackage{graphicx}
\usepackage{textcomp}
\usepackage{xcolor}
\usepackage{xspace}
\usepackage{comment}
\usepackage{kotex}
\usepackage{enumitem}
\usepackage{multirow}

\copyrightyear{2026}
\acmYear{2026}
\setcopyright{cc}
\setcctype{by}
\acmConference[DAC '26]{63rd ACM/IEEE Design Automation Conference}{July 26--29, 2026}{Long Beach, CA, USA}
\acmBooktitle{63rd ACM/IEEE Design Automation Conference (DAC '26), July 26--29, 2026, Long Beach, CA, USA}
\acmDOI{10.1145/3770743.3803941}
\acmISBN{979-8-4007-2254-7/2026/07}

\newcommand{\ArchName}{ZK-Flex\xspace}
\title{\ArchName: A Flexible and Scalable Framework for Accelerating Zero-Knowledge Proofs}

\author{Adiwena Putra}
\authornote{Both authors contributed equally to this work.}
\affiliation{%
  \institution{KAIST}
  \city{Daejeon}
  \country{South Korea}
}
\email{adiwena.research@kaist.ac.kr}

\author{Cuong Manh Duong}
\authornotemark[1]
\affiliation{%
  \institution{KAIST}
  \city{Daejeon}
  \country{South Korea}
}
\email{cuongdm1410@kaist.ac.kr}

\author{Anh Quang Pham}
\affiliation{%
  \institution{KAIST}
  \city{Daejeon}
  \country{South Korea}
}
\email{anhpq3105@kaist.ac.kr}

\author{Joo-Young Kim}
\affiliation{%
  \institution{KAIST}
  \city{Daejeon}
  \country{South Korea}
}
\email{jooyoung1203@kaist.ac.kr}

\usepackage[font=small,skip=2pt]{caption}
\setlist[itemize]{topsep=2pt, partopsep=0pt, itemsep=2pt, parsep=0pt, leftmargin=*}
\setlist[enumerate]{topsep=2pt, partopsep=0pt, itemsep=2pt, parsep=0pt, leftmargin=*}

\begin{document}

\begin{abstract}

Zero-knowledge proofs (ZKP) allows a prover to convince a verifier of computational correctness without revealing private data, ensuring both privacy and verifiability. However, proof generation is highly compute-intensive, dominated by polynomial (POLY) and elliptic-curve (EC) operations. These workloads pose two key challenges for hardware acceleration: (1) efficiently supporting diverse large-precision modular multiplications, and (2) maintaining high utilization across workloads that dynamically shift between POLY and EC stages. Existing reconfigurable accelerators address these issues only partially, remaining limited in precision scalability, algorithmic flexibility, and resource efficiency.

To overcome these limitations, we propose \ArchName, a flexible and scalable software–hardware co-designed framework for accelerating ZKP proof generation. The software layer incorporates POLY and EC optimizers that reduce computation through hardware- and workload-aware algorithmic choices, while the hardware integrates TCore, a Toom–Cook–based multi-precision core with a flexible NoC and a linked-list memory mechanism that improves parallelism under limited memory capacity. Across representative ZKP benchmarks, \ArchName achieves $5$–$11\times$ speedup and up to $3.8\times$ higher area efficiency over the state of the art, establishing a new foundation for high-performance, reconfigurable ZKP acceleration.

\end{abstract}

\ccsdesc[500]{Computer systems organization~Parallel architecture}
\ccsdesc[500]{Security and privacy~Cryptography}

\keywords{Zero-Knowledge Proof, Hardware Accelerator, Reconfigurable Architecture}

\maketitle

\section{Introduction}
\label{section1}

A Zero-Knowledge Proofs (ZKP) is a cryptographic protocol that allows a prover to convince a verifier of a statement’s correctness without revealing any secret information~\cite{zkp_ori}. The resulting proof can be verified much more efficiently than re-executing the original computation. A typical use case is outsourcing, where a trusted but resource-limited client delegates heavy computation to an untrusted yet powerful server and verifies the result through a succinct ZKP. In practice, ZKPs are widely adopted in blockchain scalability, enabling techniques such as rollups~\cite{rollup_survey}, where the blockchain acts as the slow but trusted verifier, and high-performance off-chain infrastructure performs most computations to boost throughput.

\begin{figure}[t]
    \centering
    \includegraphics[width=0.90\linewidth]{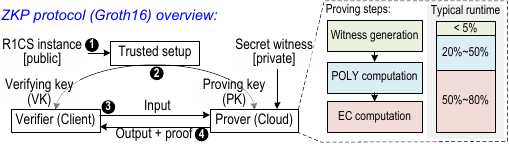}
    \caption{Overview of the Groth16 ZKP protocol.}
    \Description{}
    \label{fig:groth16}
\end{figure}

Among existing ZKP protocols, Groth16 is the most mature and widely adopted, particularly in blockchain systems, because of its compact proof size~\cite{groth16}. As shown in Figure~\ref{fig:groth16}, it involves a verifier (client), a prover (cloud), and a trusted setup that generates proving and verifying keys. The proof generation phase must be executed for every new input and dominates the overall runtime, often taking from several minutes to hours depending on application complexity~\cite{zkp_tree,zkp_vsql,zkp_zcash}. It consists of three main stages: (1) witness generation, which populates the witness vector, (2) polynomial (POLY) computation, which constructs circuit polynomials and derives the quotient polynomial $H(X)$ using the number theoretic transform (NTT), and (3) elliptic-curve (EC) computation, which encodes the witness and $H(X)$ through multi-scalar multiplication (MSM) to generate the final proof. In practice, the POLY and EC computations together account for roughly 95\% of the total proving time.

\textbf{Challenges in accelerating proof generation.} Designing a hardware accelerator for proof generation presents several challenges. First, supporting large-bitwidth modular multiplication is difficult. While common in cryptographic accelerators (e.g., FHE~\cite{abcfhe,putra2024morphling,putra2023strix,deng2024trinity,kim2023sharp}), it is more constrained in ZKP systems because computation cannot exploit Residue Number System (RNS) parallelism: Groth16 uses large prime moduli that do not decompose into smaller coprime integers, and the modulus varies across stages (e.g., 381-bit base field for EC vs. 255-bit scalar field for POLY in BLS12-381~\cite{BLS12-381}). Second, proof generation requires multiple large-precision kernels, namely NTT and MSM, executed sequentially, making balanced utilization difficult. Prior designs with separate engines for NTT and MSM~\cite{zhang2021pipezk,daftardar2024szkp} often suffer from underutilization, especially as workload size shifts the bottleneck between POLY and EC stages, leaving one engine idle for significant periods.

\textbf{Prior works}. To address this problem, a specialized hardware architecture that can be reconfigured to support various large-bitwidth precisions and compute kernels (i.e., NTT and MSM) is required. Two notable prior works that fall into this category are ReZK~\cite{zhou2024rezk} and LegoZK~\cite{yang2025legozk}. However, these designs have key limitations: ReZK supports only up to 384-bit precision, while LegoZK handles larger bitwidths but suffers from suboptimal area efficiency. In addition, both architectures use fixed algorithmic mappings for NTT butterflies and EC operations, preventing them from exploiting more resource-efficient algorithmic variants that could improve parallelism under certain workloads or cryptographic parameters. Furthermore, prior works primarily focus on hardware-only optimizations, overlooking potential software-level strategies that can significantly reduce the overall computational workload. 

\begin{figure}[t]
    \centering
    \includegraphics[width=0.90\linewidth]{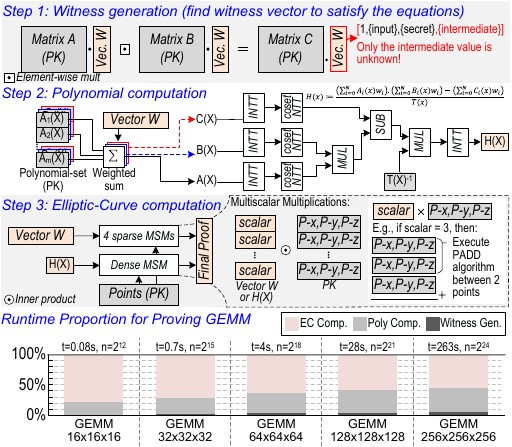}
    \caption{Groth16's proving step breakdown and workload distribution analysis using GEMM case.}
    \Description{}
    \label{fig:workload}
\end{figure}

\textbf{Our solution}. We propose \ArchName, a flexible and scalable framework that accelerates ZKP generation via software–hardware co-optimization. The software layer selects resource-efficient algorithms and tiling strategies based on workload and resource constraints, while the hardware provides a flexible arithmetic architecture with TCore, a Toom–Cook–based core for area-efficient multi-precision multiplication~\cite{toom1963complexity}. Multiple TCore units can be composed to realize various NTT butterflies and EC operations. Finally, we introduce a linked-list memory management mechanism that improves compute-engine utilization during MSM accumulation, reducing memory conflicts and enabling \ArchName\ to achieve high area efficiency across diverse ZKP workloads.
The main contributions of \ArchName are as follows:
\begin{itemize}[leftmargin=1em, topsep=2pt, itemsep=2pt, parsep=0pt]
\item We propose \ArchName, a unified framework that alleviates ZKP proof-generation bottlenecks through software–hardware co-optimization. The software layer selects workload-efficient configurations and tiling strategies, while the hardware supports diverse large-bitwidth arithmetic kernels.
\item On the software side, \ArchName introduces a \textit{Next Smooth Composite} (NSC) padding strategy for NTT computation, which reduces the number of points processed in the MSM stage. For MSM, it automatically determines the optimal window size and tiling strategy based on available memory capacity and parallelism.
\item On the hardware side, \ArchName integrates a Toom–Cook–based arithmetic core (TCore) configurable across multiple bitwidths, and introduces a linked-list memory management mechanism that improves MSM accumulation efficiency and utilization.
\item We evaluate \ArchName across diverse ZKP workloads, demonstrating $5-11\times$ speedup and up to $3.8\times$\ higher area efficiency compared with the best prior accelerator~\cite{yang2025legozk}.
\end{itemize}

\begin{figure}[t]
    \centering
    \includegraphics[width=0.99\linewidth]{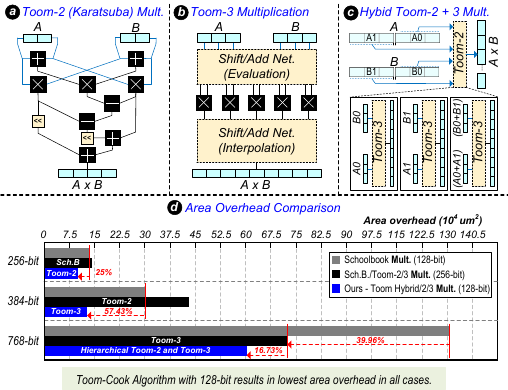}
    \caption{(a-c) Various implementations for multi-precision multiplication. (d) Hybrid Toom-2/3 achieves the best area overhead}
    \Description{}
    \label{fig:toomcook}
\end{figure}

\section{Background \& Challenges}
\label{section2}

\begin{figure*}[t]
    \centering
    \includegraphics[width=0.90\linewidth]{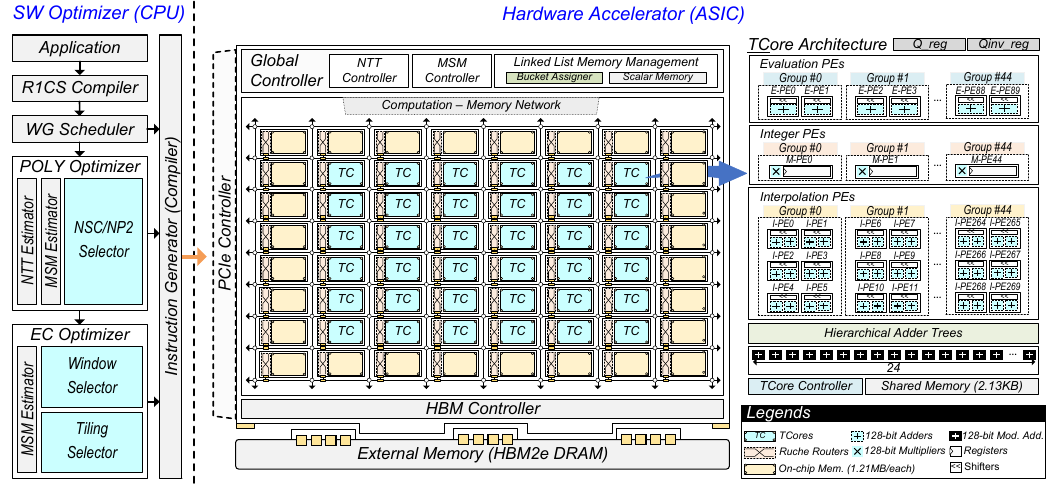}
    \caption{Overview of the \ArchName framework. The software stack on the host CPU optimizes and generates instructions for the hardware, which is an ASIC accelerator with reconfigurable cores supporting multiple bitwidths and high-level kernels.}
    \Description{}
    \label{fig:overall_arch}
\end{figure*}

\subsection{Groth16 Construct}
\textbf{Groth16 proving steps.} We refer readers to Groth16~\cite{groth16} for full protocol details and focus on the computation-intensive stages in Figure~\ref{fig:workload}. The process consists of three main steps: (1) In the witness generation (WG) stage, the prover uses an R1CS solver to compute all witness values that satisfy the constraint system. (2) In the polynomial computation stage, the witness vector is used to form the polynomials $A(X)$, $B(X)$, and $C(X)$, which are combined to obtain the quotient polynomial $H(X)$. This stage is dominated by the Number-Theoretic Transforms (NTT and INTT). (3) In the elliptic-curve (EC) computation stage, the prover encodes $H(X)$ and the witness vector through pairing-based commitments to produce the final proof. The main workload in this stage is multi-scalar multiplication (MSM), which involves point additions (PADD) and doublings (PDBL). To reduce computation and improve parallelism, we adopt the Pippenger algorithm for MSM~\cite{pippenger1976evaluation}.

\textbf{Workload analysis}. 
Figure~\ref{fig:workload} (bottom panel) illustrates proof generation time for general matrix multiplication (GEMM) across different matrix sizes on a single-thread CPU. For small number of constraints, runtime is dominated by EC operations, but polynomial computation grows rapidly with problem size and soon becomes the main bottleneck. Since GEMM reflects common computation patterns in ML workloads, where matrices often contain thousands of elements, polynomial computation is expected to match or surpass EC cost in large-scale proofs. This trend highlights the limitation of prior accelerators that focus only on EC operations~\cite{edmsm, pipemsm, priormsm, hardcaml} or use separate engines for polynomial and EC processing, leading to poor resource utilization and unbalanced performance.

\vspace{-4pt}
\subsection{Large bit-width computation challenge}
ZKP protocols operate over large finite fields, where operand widths are typically 256, 384, or 768 bits depending on the elliptic-curve parameters~\cite{yang2025legozk}. Such large-precision arithmetic requires partitioning operands into 128-bit slices, where the number of slices is denoted as $s$. The standard schoolbook multiplication then requires $s^2$ partial products and roughly $3s^2$ modular multiplications (including reduction), resulting in high latency and resource cost. Prior accelerators have employed Karatsuba multiplication to reduce complexity~\cite{karatsuba1963multiplication}, but these designs efficiently support only 256- and 384-bit operands. Extending to 768 bits requires deeper recursion or larger 256-bit slices, both of which increase latency and lower resource utilization.

To overcome these limitations, we adopt the Toom–Cook (TC) algorithm~\cite{toom1963complexity,division_free_tc}, a generalized form of Karatsuba multiplication. As shown in Figure~\ref{fig:toomcook}, we use Toom-2 for 256-bit and Toom-3 for 384-bit operations, both implemented without additional recursion. For 768-bit operands, we introduce a hybrid Toom-2/Toom-3 scheme that requires only two recursive levels. Synthesized results in 28 nm (Figure~\ref{fig:toomcook}(d)) show that this hybrid achieves the lowest area overhead compared with schoolbook, Toom-2-only, and Toom-3-only designs. The key challenge is building a compute core that flexibly supports both Toom-2 and Toom-3 modes while maintaining high area efficiency. To our knowledge, this is the first hierarchical Toom–Cook architecture applied to ZKP hardware acceleration.

\section{ZK-Flex Framework}
\label{section3}

\subsection{Framework Overview}
Figure~\ref{fig:overall_arch} shows the \ArchName framework, which integrates a host-side software optimizer with a specialized ASIC accelerator. The software optimizer consists of three modules: the witness generation (WG) scheduler, the polynomial (POLY) optimizer, and the elliptic-curve (EC) optimizer. Each module produces a sequence of hardware instructions that are executed by the accelerator. The hardware accelerator adopts a spatially distributed architecture in which compute and memory nodes are arranged in a grid and interconnected through a ruche-based network-on-chip (NoC)\cite{ruche}. Memory nodes are distributed along the mesh boundary and serve as global memory, each providing multiple SRAM banks with a total capacity of about 1.21 MB. Compute nodes, referred to as TCore (Toom–Cook cores), contain several processing elements (PEs) for performing Toom–Cook multiplications. For chip I/O, the accelerator provides two main interfaces: a PCIe interface for transferring parameters and instructions from the host, and a high-bandwidth memory (HBM) interface for data movement during computation.

\subsection{Software Optimizer} 

After compiling the application into R1CS using Circom~\cite{iden3_circom} or Gnark~\cite{gnark}, the WG scheduler executes gates level by level, enabling parallelism within each level. This strategy, also used in Gnark, is adapted in \ArchName\ for parallel hardware execution.

\begin{figure}[tbp]
    \centering
    \includegraphics[width=0.9\linewidth]{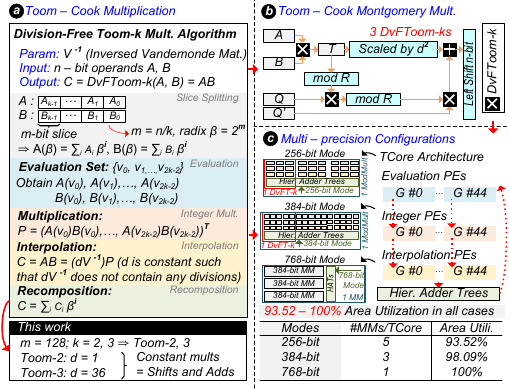}
    \caption{(a) Toom–Cook multiplication~\cite{division_free_tc}, (b) Montgomery multiplier, and (c) configurable PEs for multi-precision modmul.}
    \Description{}
    \label{fig:intra_core_config}
\end{figure}

\textbf{POLY optimizer}.
Once the witness vector is filled, it is used to construct the circuit polynomials $A(X)$, $B(X)$, and $C(X)$, which are processed through NTT/INTT to derive the quotient polynomial $H(X)$. In conventional systems (and all prior works), the NTT is optimized for radix-2 butterflies, forcing the polynomial length to be padded to the next power of two—referred to as the \textit{Next Power-of-Two (NP2)} strategy. The coefficients of $H(X)$ are then used as the scalar vector for dense MSM, so this padding directly increases the number of MSM points. When the number of constraints is far from a power of two, the workload for MSM can nearly double.

To address this inefficiency, we introduce the \textit{Next Smooth Composite (NSC)} strategy, which pads the polynomial length to the next composite number formed from a small set of prime factors $\{2,3,5,7\}$ instead of the next power of two. For example, with 33 constraints, the NP2 strategy pads to 64 ($2^6$), whereas NSC pads to 35 ($5\times7$), reducing the number of MSM points from 64 to 35. The optimizer precomputes an NSC lookup table containing smooth composite numbers to efficiently determine valid sizes. We observe a sufficient number of such composites using only two distinct prime factors. To leverage this flexibility, \ArchName incorporates a mixed-radix NTT engine that supports radix-2/3/4/5/7/8 butterflies, enabling efficient execution of NSC-padded workloads. 

A caveat is that mixed-radix NTT can be slower than radix-2 NTT, especially when the constraint size is already close to a power-of-two boundary. However, NSC may still improve end-to-end performance if reductions in MSM cost outweigh the mixed-radix NTT overhead. To balance this trade-off, the POLY optimizer includes analytical estimators for both NTT and MSM, evaluates NP2 and NSC configurations for each workload, and selects the strategy that minimizes end-to-end proving time.

\textbf{EC optimizer}. 
We adopt the Pippenger bucket-based MSM algorithm~\cite{pippenger1976evaluation}, which partitions scalars into fixed-size windows and accumulates points into buckets. The EC optimizer applies two optimizations: (1) selecting the window size that minimizes point additions and doublings using analytical models~\cite{edmsm}, and (2) determining a tiling strategy based on memory capacity and parallelism. While conventional window-level parallelism assigns each PADD engine to a different window to avoid conflicts~\cite{zhang2021pipezk, daftardar2024szkp, yang2025legozk}, it requires per-engine bucket storage that scales poorly. We instead allow multiple PADD engines to process the same window while sharing a unified bucket memory, with a linked-list–based mechanism ensuring conflict-free updates (described in next subsection).

\subsection{Hardware Accelerator}
\textbf{Supporting multiple large-bitwidth precisions}.
To support multiple large-bitwidth precisions (256, 384, and 768 bits) using the Toom–Cook algorithm, we design a reconfigurable core (\textbf{TCore}) composed of three PE types corresponding to the evaluation, integer multiplication, and interpolation stages. As shown in Figure~\ref{fig:overall_arch}, PEs of the same type are organized into 45 groups. Adjacent PE groups across different types are connected to form what we call a PE-thread. Figure~\ref{fig:intra_core_config} illustrates how the Toom–Cook algorithm maps onto our TCore architecture. Multiple PE-threads can be combined into a PE-thread group to collaboratively perform higher-bitwidth integer multiplications. For example, a single 256-bit integer multiplier is formed by combining three PE-thread groups. A Montgomery multiplication is then constructed by composing three such 256-bit integer multipliers together with the modular adder (ModAdd) unit, which is integrated within TCore (see Figure~\ref{fig:intra_core_config}). To enable efficient data transfer between different PE-thread groups and the ModAdd unit, each TCore includes a small shared memory for inter-thread communication. This hierarchical organization of PE-threads, thread groups, and shared memory enables TCore to efficiently scale across various bitwidths while maintaining high hardware utilization and modular design flexibility.

\begin{figure}[t]
    \centering
    \includegraphics[width=0.9\linewidth]{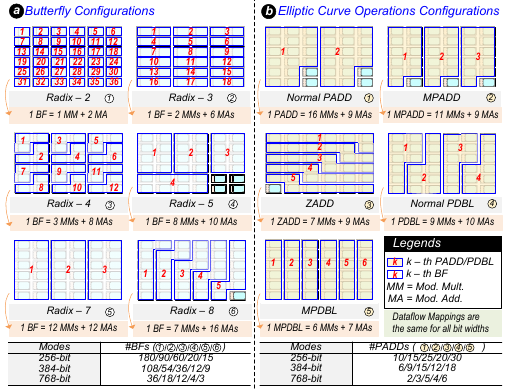}
    \caption{Composing multiple TCores via NoC to form higher-level kernels: (a) NTT butterflies and (b) EC operations.}
    \Description{}
    \label{fig:inter_core_config}
\end{figure}

\textbf{Supporting mixed-radix NTT computation}.
Multiple TCore units can be composed to implement NTT and MSM kernels (Fig.~\ref{fig:inter_core_config}). For NTT, the architecture supports six butterfly types (radix 2, 3, 4, 5, 7, 8) to match mixed-radix schedules from NSC padding. Accelerating NTT introduces two major challenges. (1) First, mixed-radix NTT requires substantial data shuffling between stages. Because memory nodes lie at the boundary of our ruche NoC, frequent cross-node reads and writes incur large network delays: each transition requires reconfiguring the NoC and refilling its pipeline before reaching steady-state throughput. To avoid reconfiguration at every stage, we adopt a constant-geometry butterfly structure, ensuring identical access patterns across all stages. (2) Second, the limited number of instantiated butterfly units and the restricted on chip memory, especially for higher radix and high security settings, prevents fully pipelined NTT execution across all stages. The total number of stages also varies with number of constraints. To address this, we tile the workload and compute the NTT one stage at a time. Butterflies within the same stage operate in parallel and share a common radix, which eliminates inter butterfly communication and simplifies NoC routing while improving efficiency.

\textbf{Supporting various elliptic-curve (EC) operations for MSM computation}.
For MSM computation, multiple TCore units can be composed to implement various point arithmetic operations from the explicit formula database~\cite{EFD}, allowing our architecture to operate as a general-purpose elliptic curve accelerator. Figure~\ref{fig:inter_core_config}(b) shows examples of point operations built from TCore compositions. This flexibility enables us to select the most efficient formulas for the MSM kernel~\cite{efdd}. 
We further apply three optimizations: (1) signed-digit representation to reduce bucket count without increasing windows~\cite{pipemsm}, (2) parallel reverse prefix sum for bucket aggregation~\cite{priormsm}, and (3) mixed point addition (MPADD), which operates on affine–Jacobian pairs and reduces modular multiplications~\cite{edmsm}, enabling more parallel engines.
However, simply increasing the number of MPADD engines does not guarantee full utilization. As discussed previously, in the conventional window-level parallelism strategy, each MPADD engine processes a distinct window and requires its own bucket memory set to avoid memory conflicts. 
If sufficient bucket memory sets are unavailable, the fully unrolled window-level strategy cannot be applied, forcing multiple MPADD engines to share the same window.

\begin{figure}[t]
    \centering
    \includegraphics[width=0.9\linewidth]{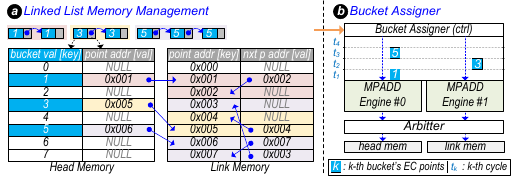}
    \caption{(a) Linked-list memory chains point addresses within each bucket. (b) Bucket assigner maps buckets to MPADD engines.}
    \Description{}
    \label{fig:linked_list}
\end{figure}

\textbf{Linked-list memory management for enhanced parallelism under limited memory space}.
To address memory constraints, we introduce a linked-list memory technique (Fig.~\ref{fig:linked_list}) that clusters points in each bucket via pointers, enabling dynamic allocation to MPADD engines with conflict-free access to bucket and point memories. 
The linked-list memory is populated on the fly as scalar information is written to the chip, thereby hiding the associated latency. This method increases the scalar memory requirement by roughly two times compared to the standard design, but the scalar memory footprint remains negligible compared to the bucket memory. The implementation requires two small memories: (1) a head memory that stores the initial pointers for each bucket, and (2) a link memory where each entry records the next point address and its corresponding linked list pointer. Finally, a bucket assigner distributes buckets across MPADD engines to ensure conflict free updates within a shared window. 
By adopting this technique, we enable highly parallel MPADD operations while preventing the superlinear growth of bucket memory as parallelism increases.

\section{Evaluation}

\subsection{Hardware Modeling}
For the hardware model of \ArchName, we implement the PEs, shared memories, TCores, and interconnects in SystemVerilog and synthesize them using a commercial 28 nm PDK, targeting minimal area and power at a 1 GHz frequency. \ArchName employs single-port, multi-bank, 128-bit, 1 GHz SRAMs for the on-chip memory nodes, providing 34 MB of total capacity. Each TCore also includes a 2.13 KB local shared memory using the same SRAM organization. TCore–TCore links use 768-bit ports at 1 GHz, yielding 4.608 TB/s of bisection bandwidth across the full-duplex $8\times8$ ruche network. Table~\ref{table:area_power_breakdown} summarizes the synthesis results for a configuration with 36 TCore and 28 memory-node instances. The complete chip occupies 205.512 mm$^2$ and consumes up to 48.520 W at 1 GHz.

\begin{table}[t]
\centering
\caption{Area and power breakdown of \ArchName.}
\label{table:area_power_breakdown}
\footnotesize
\setlength{\tabcolsep}{5.2pt} 
\renewcommand{\arraystretch}{0.72} 
\begin{tabular}{lcc}
\toprule
\textbf{Component} & \textbf{Area (mm$^2$)} & \textbf{Power (W)} \\
\midrule
45$\times$ Eval. PE Group     & 0.023  & 0.0099 \\
45$\times$ Int. PE Group.      & 1.507  & 0.7488 \\
45$\times$ Interp. PE Group.   & 0.137  & 0.0594 \\
Hier. Adder Tree             & 0.251  & 0.1090 \\
TCore Ctrl.                  & 0.004  & 0.0096 \\
24$\times$ ModAdder (128b)   & 0.069  & 0.0240 \\
Shared Mem. (2.13 KB)        & 0.0065 & 0.0006 \\
\midrule
\textbf{TCore subtotal}      & \textbf{1.998} & \textbf{0.961} \\
\midrule
36$\times$ TCore             & 71.924 & 34.998 \\
28$\times$ Mem. Node (1.21MB)& 105.756 & 10.467 \\
Global Ctrl.                 & 0.988  & 0.098  \\
Ruche NoC                    & 3.745  & 0.371  \\
HBM2e PHY                    & 23.095 & 1.907  \\
\midrule
\textbf{Total}               & \textbf{205.512} & \textbf{48.520} \\
\bottomrule
\end{tabular}
\vspace{-0.6em}
\end{table}

\begin{table*}[t]
\centering
\caption{Application benchmarks and comparison with prior works. Reported metrics include POLY/EC runtimes (ms) and area efficiency (AE = proofs/s/mm\textsuperscript{2}). Curves: MNT4-753 (AES–Auction), BLS12-381 (Sprout–Sapling Out), BN128 (Lenet–VGG16).}
\label{table:zkp_breakdown}
\scriptsize
\setlength{\tabcolsep}{3.0pt} 
\renewcommand{\arraystretch}{0.72} 
\begin{tabular}{l r|
cccc|cccc|cccc|cccc|ccc|ccc}
\toprule
\textbf{Application} & \textbf{\#Const.} &
\multicolumn{4}{c|}{\textbf{PipeZK (PZK)}} &
\multicolumn{4}{c|}{\textbf{GZKP}} &
\multicolumn{4}{c|}{\textbf{LegoZK (LZK)}} &
\multicolumn{4}{c|}{\textbf{This Work}} &
\multicolumn{3}{c|}{\textbf{Speedup (×)}} &
\multicolumn{3}{c}{\textbf{Area Eff. (×)}} \\ 
\cmidrule(lr{0pt}){3-24}
 & &
POLY & EC & Tot & AE &
POLY & EC & Tot & AE &
POLY & EC & Tot & AE &
POLY & EC & Tot & AE &
PZK & GZKP & LZK &
PZK & GZKP & LZK \\ 
\midrule
AES            & 14.2k  & 2    & 97    & 99    & 190.6  & 4    & 99   & 103   & 11.91 & 0.121 & 11   & 11.12 & 1.31k & 0.365 & 1.691 & 2.06  & 2.37k & 48.15 & 50.10 & 5.41 & 12.40 & 198.7 & 1.81 \\
SHA            & 25.7k  & 3    & 102   & 105   & 180.0  & 5    & 66   & 71    & 17.28 & 0.272 & 9    & 9.27  & 1.57k & 0.383 & 1.936 & 2.32  & 2.10k & 45.28 & 30.62 & 4.00 & 11.66 & 121.4 & 1.34 \\
RSA            & 93.7k  & 14   & 1.23k & 1.24k & 15.19  & 22   & 120  & 142   & 8.64  & 2.000 & 17   & 19.0  & 765.1 & 1.533 & 7.382 & 8.92  & 545.8 & 139.5 & 15.93 & 2.13 & 35.93 & 63.17 & 0.71 \\
RSA--SHA       & 117.7k & 14   & 822   & 836   & 22.61  & 24   & 130  & 154   & 7.96  & 2.000 & 19   & 21.0  & 692.2 & 1.761 & 8.100 & 9.86  & 493.4 & 84.78 & 15.62 & 2.13 & 21.83 & 61.93 & 0.71 \\
Merkle--Tree   & 295k   & 63   & 2.70k & 2.76k & 6.84   & 60   & 220  & 280   & 4.38  & 6.000 & 33   & 39.0  & 372.7 & 4.134 & 15.41 & 19.55 & 248.9 & 141.2 & 14.33 & 2.00 & 36.35 & 56.81 & 0.67 \\
Auction        & 541k   & 139  & 2.05k & 2.19k & 8.62   & 150  & 370  & 520   & 2.35  & 12.00 & 42   & 54.0  & 269.2 & 12.95 & 23.47 & 36.42 & 133.6 & 60.19 & 14.28 & 1.48 & 15.50 & 56.62 & 0.50 \\
Sprout         & 1.99M  & 76   & 677   & 753   & 26.93  & 49   & 250  & 299   & 4.10  & 8.000 & 19   & 27.0  & 538.4 & 6.998 & 8.849 & 15.85 & 307.1 & 47.52 & 18.87 & 1.70 & 11.40 & 74.82 & 0.57 \\
Sapling Spend  & 98.8k  & 40   & 167   & 207   & 97.99  & 3    & 90   & 93    & 13.19 & 0.515 & 7    & 7.51  & 1.93k & 0.667 & 1.611 & 2.28  & 2.14k & 90.87 & 40.83 & 3.30 & 21.80 & 161.9 & 1.10 \\
Sapling Out    & 7.83k  & 0.25 & 34    & 34.25 & 592.2  & 1    & 33   & 34    & 36.08 & 0.022 & 2    & 2.02  & 7.18k & 0.053 & 0.313 & 0.366 & 13.3k & 93.59 & 92.90 & 5.52 & 22.45 & 368.4 & 1.85 \\
Lenet          & 865k   & 20   & 1.09k & 1.10k & 18.27  & 7.49 & 0.05 & 0.05  & 23.37 & 1.000 & 60   & 61.0  & 238.3 & 4.061 & 2.227 & 6.29  & 773.8 & 176.5 & 8.35  & 9.70 & 42.35 & 33.10 & 3.25 \\
AlexNet        & 793.8M & 80.7k& 557k  & 637.8k& 0.031  & 12.27& 44.64& 56.91 & 0.022 & 2.1k  & 41.3k& 43.4k & 0.334 & 2.80k& 1.13k & 3.9k  & 1.237 & 162.2 & 14.48 & 11.1 & 38.92 & 57.42 & 3.70 \\
VGG16          & 1.42B  & 175.6k& 981k & 1.15M & 0.017  & 24.32& 89.28& 113.6 & 0.011 & 4.5k  & 83.5k& 88.1k & 0.165 & 5.63k& 2.01k & 7.6k  & 0.636 & 151.4 & 14.86 & 11.5 & 36.32 & 58.93 & 3.86 \\
\bottomrule
\end{tabular}
\vspace{-0.3em}
\end{table*}

\begin{figure*}[t]
    \centering
    \includegraphics[width=0.95\linewidth]{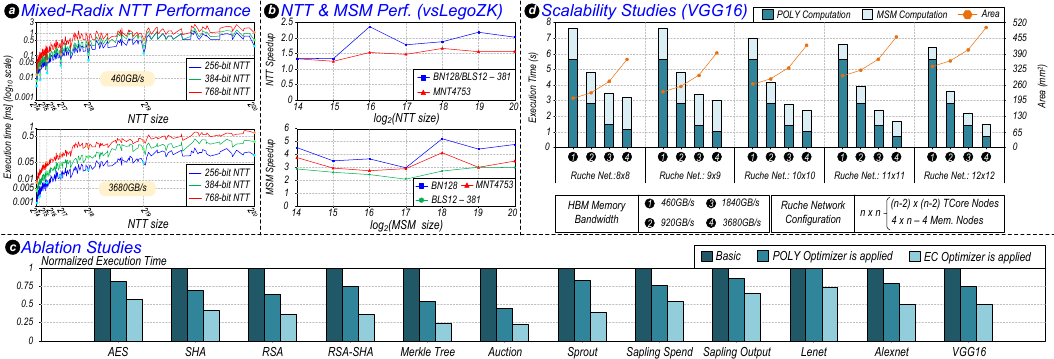}
    \caption{Evaluation of \ArchName:
    (a) Mixed-radix NTT runtime across composite sizes.
    (b) Radix-2 NTT and MSM speedup over LegoZK.
    (c) Ablation of software optimizations.
    (d) Scalability with core count and off-chip bandwidth.}
    \Description{}
    \label{fig:eval}
\end{figure*}

\subsection{Evaluation Methodology}
\textbf{Performance modeling}.
To evaluate \ArchName, we build a software optimizer and a custom cycle-level hardware simulator for the ASIC accelerator. Realistic workloads are generated using existing R1CS compilers (e.g., Gnark or Circom), enabling accurate modeling of witness-generation runtime and sparse MSM behavior. The hardware simulator incorporates synthesis results from Design Compiler for power and area estimation and uses Ramulator to model HBM latency~\cite{ramulator}. Unless stated otherwise, we assume a 1 GHz operating frequency and 460 GB/s HBM bandwidth.

\textbf{Experimental setup}.
We first evaluate the raw ASIC performance of \ArchName for NTT and MSM without enabling the software optimizer. For NTT, we benchmark the mixed radix engine and report runtimes across all supported composite sizes. We then compare both our NTT and MSM performance against the prior state of the art reconfigurable ZKP accelerator, LegoZK~\cite{yang2025legozk}. For a fair comparison, we use radix-2 NTT and evaluate MSM only at power-of-two sizes. For application level benchmarks, we enable the software optimizer and compare \ArchName with prior works~\cite{zhang2021pipezk, gzkp, yang2025legozk} under identical configurations to show maximal achievable performance. To analyze the impact of each optimization, we conduct an ablation study on the ZKML workload. Finally, we evaluate scalability by varying the number of compute cores and available off chip bandwidth to examine performance proportionality.

\subsection{Results and Discussion}

\textbf{NTT \& MSM performances}.
Figure~\ref{fig:eval}(a) shows mixed-radix NTT performance across all composite sizes formed from radix 2, 3, 5, and 7 under two bandwidth configurations. As expected, pure powers of two give the best performance, while composite sizes reduce NTT throughput by roughly 50\% on average due to memory-bandwidth bottlenecks. Higher bandwidth mitigates this slowdown, and \ArchName may still select mixed-radix sizes when the resulting reduction in MSM cost outweighs the NTT slowdown. Figure~\ref{fig:eval}(b) compares radix-2 NTT and MSM throughput against LegoZK. \ArchName achieves average NTT speedups of $1.48\times$ (MNT4-753) and $1.84\times$ (BLS12-381, BN128), and MSM speedups of $4.16\times$, $2.69\times$, and $3.29\times$ on BN128, BLS12-381, and MNT4-753, respectively. These results confirm that \ArchName delivers substantial acceleration for both NTT and MSM across curves and input sizes.

\textbf{Application benchmark}.
To assess end-to-end proving performance, we benchmark \ArchName across a wide range of real-world applications and compare against PipeZK~\cite{zhang2021pipezk}, GZKP~\cite{gzkp}, and LegoZK~\cite{yang2025legozk}, representing heterogeneous, GPU, and reconfigurable accelerators. Table~\ref{table:zkp_breakdown} summarizes the results. Compared with the best prior design (LegoZK), \ArchName achieves an average $5.52\times$ speedup in proof generation latency (up to $11.5\times$ in the best case) and improves area efficiency by $1.85\times$ on average (up to $3.8\times$). Although \ArchName shows lower area efficiency on a few applications due to underutilization of TCore at certain radix values (for example, radix 5 and 8), it consistently delivers higher overall speedup, demonstrating substantial performance gains across cryptographic, blockchain, and machine learning ZKP workloads.

\textbf{Ablation study}.
We perform ablation studies on the \ArchName software optimizations across various applications to evaluate their impact on performance. Figure~\ref{fig:eval}(c) reports normalized runtime relative to a baseline without optimizations. On average, the POLY optimizer provides a $1.4\times$ speedup, with larger gains observed when the number of constraints is far from the next power-of-two value. Adding the EC optimizer on top of the POLY optimizer yields an additional $1.7\times$ improvement on average. Overall, the full software optimizer improves performance by $2.4\times$ on average and up to $4.2\times$ in the best case among the tested applications.

\textbf{Scalability study}.
To evaluate scalability, we measure the VGG16 workload under different Ruche network sizes and memory bandwidth settings, and we report the corresponding area overhead in Figure~\ref{fig:eval}(d). The results show that additional performance gains require higher off-chip bandwidth, indicating that memory throughput becomes the bottleneck at larger scales. Performance continues to improve as the number of cores increases across all bandwidth settings, demonstrating that \ArchName can effectively exploit parallelism when adequate memory bandwidth is available.

\section{Conclusion}
ZKP proof generation is dominated by large-bitwidth polynomial and elliptic-curve computations, yet prior accelerators struggle to scale across varying precisions, workloads, and algorithms, resulting in low utilization and efficiency. \ArchName\ addresses these challenges through a unified SW–HW co-design that reduces computation and improves flexibility. The SW layer introduces workload-aware optimizations, including NSC padding and adaptive MSM configuration, while the HW integrates a multi-precision TCore and a linked-list memory mechanism to improve MSM efficiency. Together, these enable high utilization and scalable performance across diverse ZKP workloads. Experiments show $5$–$11\times$ speedup and up to $3.8\times$ higher area efficiency over prior work, establishing \ArchName\ as state of the art for reconfigurable ZKP acceleration.

\begin{acks}
This work was supported by the Institute of Information \& Communications Technology Planning \& Evaluation (IITP) through the IITP-ITRC program (IITP-2026-RS-2020-II201847), and the IITP grant (No. RS-2025-02264029, Integration and Validation of an AI Semiconductor-Based Data Center Training and Inference System), all funded by the Korea government (MSIT). The EDA tool was supported by the IC Design Education Center(IDEC), Korea.
\end{acks}

\bibliographystyle{ACM-Reference-Format}
\bibliography{references}

\end{document}